\documentclass{article}
\usepackage{spconf,amsmath,graphicx}
\usepackage[utf8]{inputenc}
\usepackage{algorithm}
\usepackage{algorithmic}
\usepackage{url}
\usepackage{cite}
\usepackage[para,online,flushleft]{threeparttable}

\title{SLOGD: Speaker LOcation Guided Deflation Approach to speech separation}
\name{Sunit Sivasankaran, Emmanuel Vincent, Dominique Fohr }
\address{Université de Lorraine, CNRS, Inria, LORIA, F-54000 Nancy, France}

\begin{document}
%
\maketitle
\begin{abstract}
    Speech separation  is the process of separating multiple speakers from an audio recording. In this work we propose to separate the sources using a Speaker LOcalization Guided Deflation (SLOGD) approach wherein we estimate the sources iteratively. In each iteration we first estimate the location of the speaker  and use it to estimate a mask corresponding to the localized speaker. The estimated source is removed from the mixture before estimating the location and mask of the next source.  Experiments are conducted on a reverberated, noisy multichannel version of the well-studied WSJ-2MIX dataset using word error rate (WER) as a metric. The proposed method achieves a WER of $44.2$ \%, a $34$\% relative improvement over the system without separation and $17$\% relative improvement over  Conv-TasNet.

\end{abstract}
\begin{keywords}
Speech separation, deflation, localization
\end{keywords}
\section{Introduction}
\label{sec:intro}

The performance of automatic speech recognition (ASR) system degrades in the presence of overlapping speech. To address this issue, the speech signals must be separated before feeding them to the ASR system. 
A body of work on monaural speech separation has recently been published using time-frequency clustering \cite{hershey_deep_2016-1,kinoshita_listening_2018,kolbaek_multitalker_2017-1} or the raw waveform \cite{luo_conv-tasnet_2019}. These results were however reported in clean conditions without the influence of noise and reverberation which are known to impact the speech separation performance.
 
 The problem  of speech separation has also been studied in the multichannel setting. Computational auditory scene analysis \cite{wang_computational_2006-1}  based systems cluster time-frequency bins dominated by the same source using cues such as pitch and interaural time and level differences. In \cite{wang_combining_2019},  a neural network is trained  using the phase differences of the multichannel short-time Fourier transform (STFT) as input features to learn such cues. Deep clustering \cite{hershey_deep_2016-1} is then used to associate the cues to the right source. This method generalizes well to unseen speakers but clustering faces inherent limitations such as estimating the number of clusters \cite{higuchi_deep_2017} and choosing an appropriate clustering algorithm to optimally model the embedding space \cite{kinoshita_listening_2018}. The studies in \cite{perotin_multichannel_2018-2,chen_multi-channel_2018} propose a spatial location-based approach instead, where the direction of arrival (DOA) of each speaker is assumed to be known and is used to beamform the multichannel signal in that direction. The beamformed signal is input to a neural network that estimates a time-frequency mask corresponding to the speaker. The beamformed signal contains enhanced components of the desired signal which allow the network to estimate the relevant mask. In our recent work \cite{sivasankaran_analyzing_2020}, we found that the signal-to-interference ratio (SIR) has a big impact on the separation performance even when the speakers are well separated in space: lower SIR makes it harder for the network to estimate the mask even when the true DOA is known. Localization errors also have an impact on the separation performance, albeit smaller \cite{sivasankaran_analyzing_2020}.
 
 To overcome these limitations, we propose to iteratively estimate the sources using a Speaker LOcalization Guided Deflation (SLOGD) approach. The concept of deflation was introduced in blind source separation and refers to the iterative estimation and removal of one source at a time \cite{delfosse_deflation_1995}. Our intuition is that the dominant sources are easier to estimate in the first few iterations and, once they have been removed from the mixture, it becomes easier to estimate the other sources. In order to implement this approach in a modern deep learning based multichannel signal processing framework, we estimate the location of a first speaker, compute the beamformed signal, and use features of that signal as inputs to a neural network to derive the corresponding mask. The location and the mask of the next source are estimated by removing the first speaker from the mixture's features using the first estimated mask, and so on. The mask estimation network is trained on beamformed signals computed from potentially erroneous DOA estimates, which results in increased robustness to localization errors.
 
 A few earlier works have proposed to iteratively estimate the sources using deep neural networks in a single-channel setting \cite{kinoshita_listening_2018,takahashi_recursive_2019}. To the best of our knowledge, this paper is the first such study in a multichannel setting, where we estimate both the DOAs and the masks of all speakers. The estimation of DOA is crucial since location-based speech separation is the only method which works well in the presence of reverberation and noise as shown in \cite{chen_multi-channel_2018}.
 
 The rest of the paper is organized as follows. Section \ref{sec:method} introduces the proposed method. Section \ref{sec:exp} describes the dataset and the experimental settings. The results are discussed in Section \ref{sec:res} and we conclude in Section \ref{sec:conclude}.

\section{Proposed method}
\label{sec:method}

\begin{figure*}
    \centering
    \includegraphics[scale=0.7]{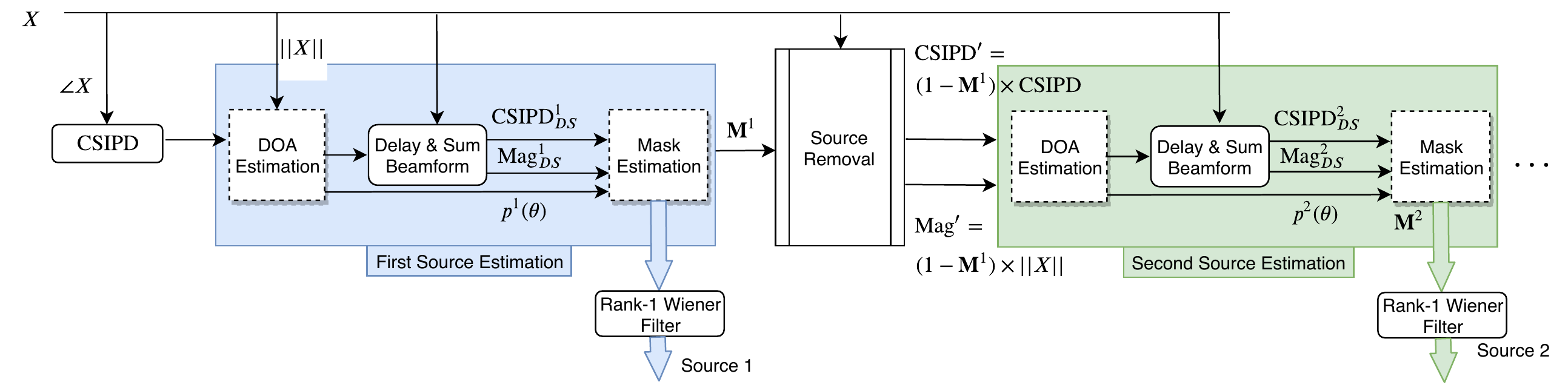}
    \caption{Iterative estimation of DOA and mask}
    \label{fig:loop_est}
\end{figure*}{}
\subsection{Signal model}
The multichannel signal $\mathbf{x}(t) = [x_1(t),  \dots, x_I(t)]^T$ captured by an array of $I$ microphones can be expressed as
\begin{equation}
    \mathbf{x}(t) = \sum_{j=1}^{J} \mathbf{c}_j(t)
\end{equation}
where $\mathbf{c}_j(t) = [c_{1j}(t), \dots, c_{Ij}(t)]^T$ is the spatial image of source $j$, i.e., the signal emitted by the source and captured at the microphones. Similarly to \cite{gannot_consolidated_2017-2}, the  microphone index and the time index are denoted by $i$ and $t$, respectively, and $J$ is the total number of sources. This general formulation is valid for both point sources as well as diffuse noise. For point sources such as human speakers, the spatial image can be expressed as a linear convolution of the room impulse response (RIR) $\mathbf{a}_j(t, \tau) = [a_{1j}(t,\tau), \dots, a_{Ij}(t, \tau)]^T$ and a single-channel source signal $s_j(t)$ as
\begin{equation}
    \mathbf{c}_j(t) = \sum_{\tau=0}^{\infty} \mathbf{a}_j( t, \tau) s_j(t-\tau).
\end{equation}{}
Under the narrowband approximation, $\mathbf{c}_j$ in the time-frequency domain can be written as
\begin{equation}
    \mathbf{c}_j(t,f) = \mathbf{a}_j(t,f) s_j(t,f)
\end{equation}
If the RIR is time-invariant, that is if the speakers and microphones are not moving, $\mathbf{a}_j(t,f) = \mathbf{a}_j(f)$.

\subsection{Estimating the sources}
Our objective is to estimate the spatial image of each of the $J$ sources. We do it iteratively as shown in Fig.~\ref{fig:loop_est} and detailed below.

    \emph{Step 1, Estimating the first DOA:} In the first step we estimate the DOA of a first speaker using a neural network. The cosines and sines of the phase differences between all pairs of microphones \cite{sivasankaran_keyword-based_2018-1,wang_combining_2019}, called 
    cosine-sine interchannel phase difference (CSIPD) features, and the short-term magnitude spectrum of one of the channels (in the following, channel 1) are used as input features (see Section \ref{subsec:archi}):
    \begin{equation}
        p^1(\theta) = \text{DOA\_DNN}_1([\text{CSIPD}, |X_1|])
    \end{equation}
    The network is trained as a classifier to estimate the DOA within a discrete grid of DOAs in every time frame. A non-speech class is also included at the output so that the network can classify a frame as non-speech if no speaker is active in that particular frame, thereby providing voice-activity detection (VAD) information. VAD labels obtained from ASR alignments are used as targets while training the network. The frame-level DOA probabilities output by the network are averaged across time, and the DOA corresponding to the highest average probability is used as the first speaker DOA. 

    \emph{Step 2, Estimating the first mask:} Given the estimated DOA, we enhance the signal by simple delay-and-sum beamforming (DS) as detailed in \cite{sivasankaran_analyzing_2020}. The magnitude spectrum of the beamformed signal ($\text{Mag}^1_\text{DS}$), its phase difference with respect to the first channel ($\text{CSIPD}^1_\text{DS}$) and the output of the DOA network ($p^1(\theta)$) are concatenated and fed to a second neural network which estimates the time-frequency mask $M_1$ corresponding to the first speaker:
    \begin{equation}
        M_1 = \text{MASK\_DNN}_1([\text{Mag}^1_\text{DS}, \text{CSIPD}^1_\text{DS}, p^1(\theta)]).
    \end{equation}
    \emph{Step 3, Removing the estimated source:} There are multiple ways to remove the  estimated source from the mixture. In \cite{kinoshita_listening_2018}, a remainder mask was computed after each iteration and appended to the network inputs to estimate the next source. In this work we use a similar idea wherein we compute a remainder mask  $(1-M_1)$  but instead of appending it to the network inputs we multiply the CSIPD and magnitude spectrum features with the remainder mask before feeding them as input to the following DOA estimation and mask estimation stages. Indeed, mask multiplication was shown to perform better than mask concatenation for speaker localization \cite{sivasankaran_keyword-based_2018-1}. 

    \emph{Step 4, Estimating the second DOA:}
    Similarly to Step 1, we estimate the DOA of the second speaker by poooling over time the frame-level DOA probabilities
    \begin{equation*}
        p^2(\theta) = \text{DOA\_DNN}_2([(1-M_1) \times \text{CSIPD}, (1-M_1)\times |X_1|]).
    \end{equation*}

    \emph{Step 5, Estimating the second mask:}
    Similarly to Step 3, we apply DS beamforming using the estimated DOA and derive the mask for the second speaker as
    \begin{equation}
    \begin{split}
        \bar{M}_2 = \text{MASK\_DNN}_2([(1-M_1)\times \text{Mag}^2_{DS}, \\ (1-M_1)\times \text{CSIPD}^2_{DS}, p^2(\theta)]).
    \end{split}
    \end{equation}
    Since this mask applies to $(1-M_1)\times\mathbf{x}$, the equivalent mask to be applied to the original signal $\mathbf{x}$ is $M_2=\bar{M}_2 \times (1-M_1)$.

The proposed method can in theory estimate DOAs and masks for any number $J$ of sources using a source counting method \cite{kinoshita_listening_2018}. In the following, we assume $J=2$. 

\subsection{Adaptive beamforming to extract the source signals}
The obtained masks, namely $M_1$ for the first speaker, $M_2$ for the second speaker, and $M_3=1-M_1-M_2$ for the noise are used to estimate the speech signals by adaptive beamforming.
For a given speaker $j\in\{1,2\}$, the covariance matrix of that speaker is first estimated as
\begin{equation}
    \mathbf{\Sigma}_j(t,f) = \alpha \mathbf{\Sigma}_j(t-1,f) + (1-\alpha) M_j(t,f) \mathbf{x}(t,f) \mathbf{x}^H(t,f)
\end{equation}
where $\alpha$ is a forgetting factor and $^H$ denotes Hermitian transposition.  Similarly, the noise covariance matrix $\mathbf{\Sigma}_n$, which includes the statistics corresponding to the other speaker and background noise, can be estimated as
\begin{multline}
    \mathbf{\Sigma}_\mathbf{n}(t,f) = \alpha \mathbf{\Sigma}_\mathbf{n}(t-1,f)\\ + (1-\alpha) (1-M_j(t,f)) \mathbf{x}(t,f) \mathbf{x}^H(t,f).
\end{multline}

The first channel of $\mathbf{c}_j(t,f)$ is then estimated via the rank-1 constrained multichannel Wiener filter (R1-MWF)  \cite{wang_rank-1_2018-1}, which imposes a rank-1 constraint on $\mathbf{\Sigma}_j(t,f)$:
\begin{equation}
    W_\text{R1-MWF}(t,f) = \frac{\mathbf{\Sigma_\mathbf{n}}^{-1}(t,f) \mathbf{\Sigma}_{R1}(t,f)}{\mu + \lambda(t,f)}\mathbf{u}_1
\end{equation}
where $\mathbf{\Sigma}_{R1}(t,f)=\sigma_j(t,f)\mathbf{h}_j(t,f)\mathbf{h}_j(t,f)^H$, $\mathbf{h}_j(t,f)$ is the principal eigenvector of $\mathbf{\Sigma_n^{-1}}(t,f) \mathbf{\Sigma}_j(t,f)$, $\sigma_j(t,f)=\text{tr}\{\mathbf{\Sigma}_j(t,f)\}/\|\mathbf{h}_j(t,f)\|^2$, $\lambda(t,f) = \text{tr}\{\mathbf{\Sigma}_\mathbf{n}^{-1}(t,f) \mathbf{\Sigma}_{R1}(t,f)\}$. We chose this beamformer due to its higher performance compared to other popular alternatives  \cite{wang_rank-1_2018-1,perotin_multichannel_2018-2,sivasankaran_analyzing_2020}.

\section{Experimental Settings}
\label{sec:exp}

\subsection{Dataset}
Experiments are conducted on the multichannel, reverberated, noisy version of the WSJ-2MIX dataset \cite{hershey_deep_2016-1} introduced in \cite{sivasankaran_analyzing_2020}. Each mixture was generated by convolving two clean Wall Street Journal (WSJ) utterances with room impulse responses (RIRs) simulated for a 4-channel Kinect-like microphone array and adding real CHiME-5 \cite{barker_fifth_2018} noise. Strong reverberation and real noise make this data quite realistic and challenging. The noise segments were randomly drawn using the VAD labels in \cite{ryant_second_2019}. The RIRs were simulated using RIR-generator \cite{habets_rir-generator_2018}. For each mixture, we generated a random room with dimensions in the range of $[3-9]$~m and a random RT60 in the range of $[0.3 - 1.0]$~s. Two speakers were placed randomly in the room at a distance of $[0.5-5.5]$~m from the microphone array. The signal-to-interference ratio between the speakers was kept at the same level as the original WSJ-2MIX dataset. Noise was added at a signal to noise ratio (SNR) of $[0-10]$ dB with respect to the first speaker. The `maximum' version of the WSJ-2MIX dataset was used wherein the length of the mixed signal is equal to the maximum of the length of the two speakers. This version addresses some of the issues raised in \cite{menne_analysis_2019} regarding the `minimum' version of the dataset which does not contain mixture signals with single-speaker segments. The simulated dataset contains $30,10, 5$ hours for train, dev and test respectively. The speech and noise signals used in train, dev and test come from different speakers and different rooms.

\subsection{Features and network architectures}
\label{subsec:archi}
The STFT was computed using $50$~ms windows with $25$~ms shift, resulting in $801$ frequency bins. The $4$ microphone channels result in $6$ microphone pairs. Since the CSIPD features consist of the sines and cosines of the phase differences between all pairs of microphones, $12$ CSIPD feature vectors, each of dimension $801$ were obtained for every time frame.

\emph{DOA estimation networks:}
The DOA estimation network (for both sources) contain a 2D convolutional neural network (CNN) which takes the $12$ CSIPD channels as inputs and throws out a single-channel output using a $5\times5$ filter. This is followed by a rectified linear unit (ReLU) nonlinearity, a dropout layer, and a max pooling layer of kernel size $2\times1$ along the frequency dimension to obtain a phase projection. Similarly, another 2D CNN is used to obtain a magnitude projection with $|X_1|$ as input. The magnitude and phase projections are concatenated and fed to a bidirectional long short-term  memory (BLSTM) layer followed by a linear layer and a sigmoid nonlinearity. The DOA space is divided into $181$ discrete angles as in \cite{sivasankaran_keyword-based_2018-1}. Together with the non-speech class, this results in $181+1$ output classes. Since multiple classes are active in every time frame due to the presence of several speakers, binary cross entropy (BCE) was used as the training cost for $\text{DOA\_DNN}_1$. After source removal, only a single active speaker remains, hence we used cross entropy as the cost function for $\text{DOA\_DNN}_2$. 

\emph{Mask estimation networks}: The mask estimation networks are 2-layer BLSTMs. The true reverberated mask corresponding to the localized speaker was used as the training target, with mean square error (MSE) as the cost function. We trained the four networks one after another in the following order: $\text{DOA\_DNN}_1 \Rightarrow $  $\text{MASK\_DNN}_1 \Rightarrow$         $\text{DOA\_DNN}_2 \Rightarrow $   $\text{MASK\_DNN}_2$. We also tried to train the DOA and mask networks jointly which yielded poor results. All networks were trained using Adam as the optimizer. 

In order to evaluate the impact of localization errors, we also trained a 2-layer BLSTM mask estimation network using the true speaker DOAs. Only the $\text{CSIPD}^1_{DS}$ and $\text{Mag}^1_{DS}$ features obtained after DS beamforming were used, since $p^j(\theta)$ is irrelevant in this case. We also evaluate the performance of this network (without retraining) using the DOAs estimated via the classical generalized cross-correlation with phase transform (GCC-PHAT) method \cite{knapp_generalized_1976}. Finally, we trained the single-channel separation method Conv-TasNet \cite{luo_conv-tasnet_2019}\footnote{\url{https://github.com/kaituoxu/Conv-TasNet}} on our dataset for comparison. 

\subsection{ASR}
Separation performance was evaluated in terms of the word error rate (WER) on the R1-MWF output. No dereverberation was performed. The acoustic model (AM) for the ASR system was trained in matched conditions: every separation method was applied to the training set and the ASR system was trained on the estimated source signals using accurate senone alignments obtained from the underlying clean single-speaker utterances.  The AM was a sub-sampled 15-layered time-delayed neural network trained using lattice-free maximum mutual information \cite{povey_purely_2016}. $40$-dimensional Mel frequency cepstral coefficients along with 100-dimensional i-vectors were used as features. A 3-gram language model was used while decoding. 

\section{Results}
\label{sec:res}
\begin{table}[]
\centering
    \caption{Baseline WER(\%) results.}
    \label{tab:baseline}
\begin{tabular}{l||l|l|l}
\hline 
Input    & \textbf{\begin{tabular}[c]{@{}l@{}}Single\\ Speaker\end{tabular}} & \textbf{\begin{tabular}[c]{@{}l@{}}Single \\ speaker + noise\end{tabular}} & \textbf{\begin{tabular}[c]{@{}l@{}}2 speakers +\\  noise\end{tabular}} \\ \hline \hline
WER & 12.5                                                             & 25.5                                                                      & 66.5                                                                  \\ \hline
\end{tabular}
\end{table}

The baseline results are shown in Table \ref{tab:baseline}. A WER of $12.5$\% was obtained for a single speaker without noise. The WER degrades to $25.5$\% with noise, showing the difficulty of the real noise for ASR. With an additional overlapping speaker, we observe a substantial WER degradation as expected.

Table \ref{tab:results} shows the ASR performance after speech separation. We obtain a WER of $35.0$\% using the true speaker DOAs, a $47$ \% relative improvement with respect to the system without speech separation, showing the positive impact of localization information on speech separation. This is the best WER that can possibly be achieved using location-guided speech separation on this dataset. The WER drops to $54.5$ \% when using the DOAs estimated by GCC-PHAT, indicating that the localization errors have an adverse impact on the performance. We obtain a WER of $44.2$ \% using our proposed method, that is a $34$\% relative improvement over the baseline without separation and a $19$\% relative improvement over GCC-PHAT based DOA estimation. In comparison,  Conv-TasNet  \cite{luo_conv-tasnet_2019} gave a WER of $53.2$ \% on our dataset.

\begin{table}[]
    \caption{WER(\%) results after speech separation}
    \label{tab:results}
\centering
\begin{threeparttable}
\begin{tabular}{l||l}
\hline
Method   & WER \\ \hline \hline
\textbf{True DOA} & 35.0        \\ \hline
    \textbf{Est DOA with GCC-PHAT}\tnote{2} & 54.5        \\ \hline
\textbf{Proposed method} & 44.2        \\ \hline
\textbf{Conv-TasNet}   & 53.2        \\ \hline
\end{tabular}
\begin{tablenotes}
\item[2] AM and separation model corresponding to the True DOA was used
\end{tablenotes}
\end{threeparttable}
\end{table}

On analyzing the errors made by the proposed SLOGD approach, we found that, for around $7$\% of the test dataset, the masks estimated in the two iterations were both closer (in terms of MSE with respect to the true masks) to the same speaker rather than two distinct speakers as expected. This suggests potential for further improvement either in the first mask estimation step or in the source removal stage. 

\section{Conclusion}
\label{sec:conclude}
In this work we approached the problem of distant speech separation in challenging noise and reverberation conditions using a deflation based strategy. For each iteration, we train a network to estimate the location of the speaker and use it to estimate a time-frequency mask corresponding to the speaker. The estimated mask is used along with a rank-1 constrained MWF to extract the signal. The estimated source is then removed by masking the signal features before extracting the next source. Using this approach, we obtain a WER of $44.2$ \% compared to the WER of $53.2$ \%  obtained by Conv-TasNet. Although the proposed method gave large improvements, the problem remains very difficult. Better localization systems and improved strategies for source removal will be studied to improve the source separation performance.

\section{Acknowledgement}
This work was made with the support of the French National Research Agency, in the framework of the project VOCADOM “Robust voice command adapted to the user and to the context for AAL” (ANR-16-CE33-0006). Experiments presented in this paper were carried out using the Grid’5000 testbed, supported by a scientific interest group hosted by Inria and including CNRS, RENATER and several Universities as well as other organizations (see https://www.grid5000.fr). The research reported here was conducted at the 2019 Frederick Jelinek Memorial Summer Workshop on Speech and Language Technologies, hosted at L'\'Ecole de Technologie Sup\'erieure (Montreal, Canada) and sponsored by Johns Hopkins University with unrestricted gifts from Amazon, Facebook, Google, and Microsoft.
\newpage
\bibliographystyle{IEEEbib}
\bibliography{ref}

\end{document}